\documentclass[pdflatex,sn-mathphys-ay,iicol]{sn-jnl}

\usepackage{pstool}
\usepackage{graphicx} 
\usepackage{subcaption}
\usepackage{amsmath,amssymb,amsfonts,amsbsy,bbm,bm}
\usepackage{stmaryrd}
\usepackage{array} 
\usepackage{multirow} 

\newtheorem{remark}{Remark}%

\newcommand{\tensor}[1]{\mbox{\boldmath ${#1}$}}
\newcommand{\IntB}{\int_{\mathcal{B}}\ }
\newcommand{\IntSB}{\int_{\partial\mathcal{B}}\ }

\theoremstyle{thmstyleone}%
\theoremstyle{thmstyletwo}%

\theoremstyle{thmstylethree}%

\begin{document}
	
\title[Article Title]{A novel multi-thickness topology optimization method for balancing structural performance and manufacturability}


\author*[1]{\fnm{Gabriel} \sur{Stankiewicz}}\email{gabriel.stankiewicz@fau.de}
\author[1]{\fnm{Chaitanya} \sur{Dev}}\email{chaitanya.dev@fau.de}
\author[1]{\fnm{Paul} \sur{Steinmann}}\email{paul.steinmann@fau.de}

\affil*[1]{\orgdiv{Institute of Applied Mechanics}, \orgname{Friedrich-Alexander-Universit\"at Erlangen-N\"urnberg}, \orgaddress{\street{Egerlandstr. 5}, \city{Erlangen}, \postcode{91058}, \state{Bavaria}, \country{Germany}}}


\abstract{Topology optimization (TO) in two dimensions often presents a trade-off between structural performance and manufacturability, with unpenalized (variable-thickness) methods yielding superior but complex designs, and penalized (SIMP) methods producing simpler, truss-like structures with compromised performance. This paper introduces a multi-thickness, density-based topology optimization method designed to bridge this gap. The proposed approach guides the design towards a predefined set of discrete, allowable thicknesses by employing a novel multilevel penalization scheme and a multilevel smoothed Heaviside projection. A continuation strategy for the penalization and projection parameters, combined with an adaptive mesh refinement technique, ensures robust convergence and high-resolution geometric features. The method is validated on standard cantilever and MBB beam benchmarks. Results demonstrate that as the number of allowable thicknesses increases, the designs systematically transition from conventional truss-like structures to high-performance, sheet-like structures. Notably, designs with as few as three discrete thickness levels achieve compliance values within 2\% of those from fully unpenalized, variable-thickness optimization, while significantly outperforming standard SIMP results. The method inherently eliminates impractically thin regions and features, both in the out-of-plane and in-plane directions and produces designs well-suited for both additive manufacturing and conventional fabrication using standard-thickness stock materials, thus maximizing both performance and manufacturability.}




\maketitle

\section{Introduction}

From the optimal truss structures of Michell \cite{michell1904lviii} to contemporary computational approaches, the pursuit of greater design freedom and practical applicability has led to many substantial developments in topology optimization (TO) in recent decades. The primary objective of TO extends beyond the search for local optima. It is a complex design tool capable of generating manufacturing-ready structures for a variety of applications. Various theoretical frameworks have emerged, including homogenization \cite{bendsoe1988generating}, density \cite{bendsoe1989optimal}, or level-set methods \cite{wang2003level}. Density-based TO, specifically employing the solid isotropic material with penalization (SIMP) approach, has gathered considerable popularity due to its straightforward formulation and abundance of educational material, and is even integrated into the majority of commercial FEM tools. The underlying goal of the SIMP formulation, i.e., $E(\rho) = \rho^p E_0$, is to "penalize" intermediate (grey) densities with $p > 1$ (usually $p = 3$), steering the optimizer towards designs primarily consisting of the solid phase ($\rho = 1$) or the void ($\rho = 0$). Consequently, in the context of SIMP, the formation of non-physical structures with intermediate densities was avoided. 

In contrast, unpenalized TO, realized for instance by setting the penalization factor to $p = 1$, fundamentally alters the design concept. Since the presence of intermediate densities does not incur an additional cost, the practical design space is significantly larger. Understanding the design variables as thicknesses (as in variable thickness optimization \cite{rossow1973finite}) rather than interpolated material densities is a suitable interpretation, making all admissible values in the range $\rho \in \left[0, 1\right]$ physically meaningful.

The unpenalized formulation offers significantly more freedom in exploring solutions tailored to performance objectives and constraints, potentially resulting in structures demonstrably superior to the discrete, black-and-white variants. A thorough study was conducted in \cite{sigmund2016non}, where penalized approaches that render Michell-type (truss-like) structures were qualitatively compared with unrestricted methods. Superior structures in terms of stiffness were obtained where thin sheets or closed-walled features were permitted, as opposed to forcefully perforated plates or truss-like features. Indeed, the thin-walled structures demonstrated in the 3D cantilever numerical study are closely related to variable-thickness sheets.

The advantages of the unpenalized density-based approach are thoroughly demonstrated in \cite{kandemir2018topology}, where variable-thickness approaches were employed to design so-called 2.5D parts. Subsequently, an extension to the SIMP method, the solid isotropic with thickness penalization (SIMTP), was proposed in \cite{yarlagadda2022solid}, which utilizes a 2.5D element with a nodal thickness variable to enable thickness variations in the context of topology optimization.

The variable-thickness method inherently enables very thin planar features, which are susceptible to buckling and generally challenging for manufacturing. A number of works have focused on eliminating very thin features within variable-thickness methods. The work of \cite{giele2021approaches} demonstrates two approaches to eliminating thin features, namely by employing two auxiliary fields inspired by the cut element method and the density approach itself, respectively. In \cite{pozo2023minimum}, various penalization rules to suppress thin features in 2.5D topology optimization were proposed. In \cite{endress2023designing}, the SIMP rule is applied to densities below a predefined threshold, preventing the formation of thin sheets. The aforementioned works present robust, manufacturing-ready structures free from unstable thin sheets.

The resurgence of unpenalized TO is, to a great extent, driven by the development of dehomogenization techniques, which utilize advanced postprocessing methods to generate Michell-type structures from unpenalized density fields \cite{groen2018homogenization}. Besides density, information about orientation is retained by employing homogenization-based topology optimization. Alternatively, truss-like structures, visually similar to those obtained by dehomogenization techniques, were retrieved from homogenization-based TO in \cite{larsen2018optimal}. A related concept was developed in \cite{li2018topology}, where an optimization method in a level-set framework was proposed in which multi-patch microstructures are obtained in place of the "intermediate" regions. Beyond unpenalized TO, variable-thickness optimization is often applied to plate and shell structures \cite{zhao2017stress, meng2022shape}, multimaterial plates \cite{banh2019topology, nguyen2022multi}, and composite laminates \cite{stegmann2005discrete, sorensen2014dmto}. Advanced techniques, like coupled thickness, shape, and topology optimization, were demonstrated in \cite{meng2022shape}, or coupled thickness and material optimization in \cite{sjolund2018new, kashanian2021novel}.

In this article, we propose a technique to design structures consisting of a number of discrete, allowable thicknesses in the context of density-based topology optimization. Our approach derives from variable-thickness sheet topology optimization, in which the final thicknesses (densities) are forced towards predefined target values. Although penalization techniques are employed to achieve the target thicknesses, we prove that the resulting structures closely match those with unrestricted thickness in terms of the performance objective. Moreover, the commonly addressed issue of very thin, buckling-prone, near-zero thickness regions (\cite{giele2021approaches, pozo2023minimum, endress2023designing}), is automatically eliminated, since the lowest non-zero density corresponds to the lowest predefined target density. 

The introduction of discrete, allowable thicknesses is motivated by economic aspects of manufacturing. The multi-thickness approach facilitates the use of conventional fabrication methods that utilize standard-thickness stock materials (e.g., acrylic, sheet metal) for cutting profiles. Subsequently, established techniques like bonding, fastening, or welding can be employed for sheet joining. In the context of additive manufacturing (AM), faster print times and more efficient material deposition can be obtained when the structure is dominated by large, flat regions. In particular, the slicing technique, in which a complex geometry is decomposed into thin, flat, printable layers (generating the G-code \cite{zhou2024additive}), can be significantly simplified by considering the proposed multi-thickness approach. In general, a reduction in geometrical complexity to a set of predefined thicknesses contributes to improved manufacturability and cost-effectiveness. 

The article is organized as follows. Section \ref{sec:topopt} discusses aspects of penalized (SIMP) and unpenalized (variable-thickness) TO. In Section \ref{sec:method}, the multi-thickness method for density-based TO is introduced. The following developments comprise the proposed multi-thickness method: the localized density penalization, the multilevel smoothed Heaviside projection filter, and the necessary continuation strategy for the local penalization and Heaviside projection sharpness. In addition, a specialized adaptive mesh strategy is employed to guarantee improved geometrical resolution and reduce computational cost. Subsequently, the common benchmark examples, the cantilever and MBB beams, are numerically tested and thoroughly analyzed with respect to the performance objective in Section \ref{sec:tests}. Further, in Section \ref{sec:am} we show additively manufactured structures using the multi-thickness and variable-thickness method and compare them in the context of geometrical quality and manufacturing times.

\section{Variable-thickness (and) topology optimization}\label{sec:topopt}

The variable-thickness sheet problem in its original form directly deals with elemental thicknesses $\tau_i$ in a 2D finite element problem \cite{rossow1973finite}. The thickness variable is bound by lower and upper limits, hence not permitting the formation of topological changes such as holes.
In the context of the density method in 2D, the differentiation between variable-thickness optimization and conventionally understood topology optimization boils down to the penalization factor $p$ in the (modified) SIMP formulation:

\begin{equation}
E(\rho_e) = \left[\rho^p\left[1 - \rho_{\rm min}\right] + \rho_{\rm min}\right]E_{0}, 
\label{eq:simp}
\end{equation}

where $E(\rho_e)$ is the effective Young's modulus in the element $e$, $\rho_{\rm min} = 10^{-9}$ and $E_{0}$ the Young's modulus of the base material. By choosing $p = 1$, formation of intermediate densities, $\rho_e \in \left(0,1\right)$, is no longer restricted. In such a case, the physically correct interpretation of intermediate densities $\rho_e$ in 2D is to consider them as the out-of-plane thickness of an extruded structure, similar to $\tau_i$ in \cite{rossow1973finite}.

\begin{figure}[h!]
	\centering
	\begin{subfigure}{82mm}
		\includegraphics[width=\linewidth]{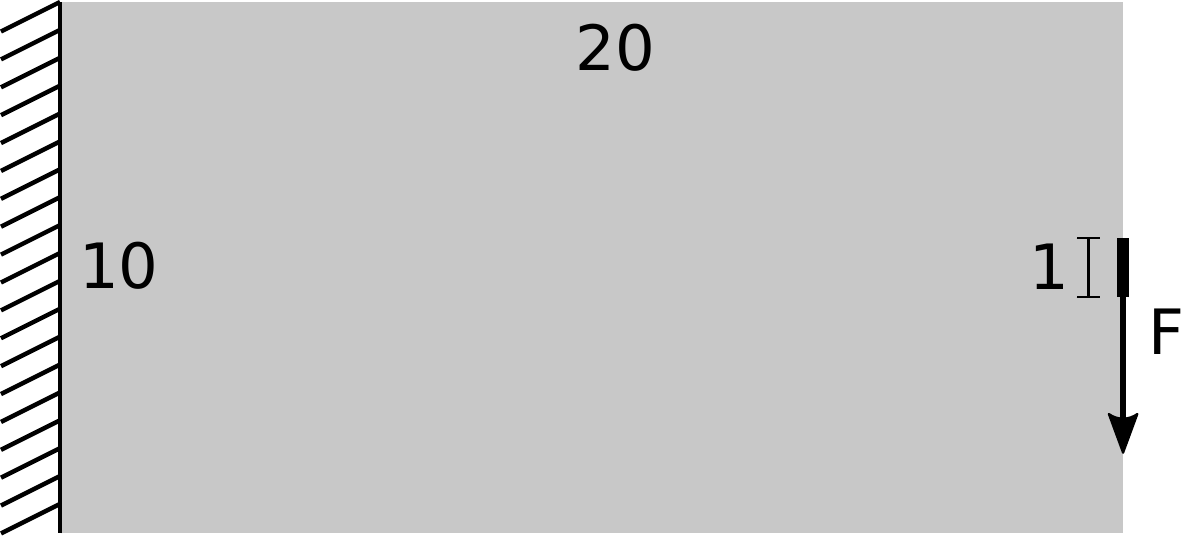}
		\caption{Cantilever beam.}
		\label{fig:first_sub}
	\end{subfigure}
	\vspace{0.02\textwidth}
	\begin{subfigure}{82mm}
		\includegraphics[width=\linewidth]{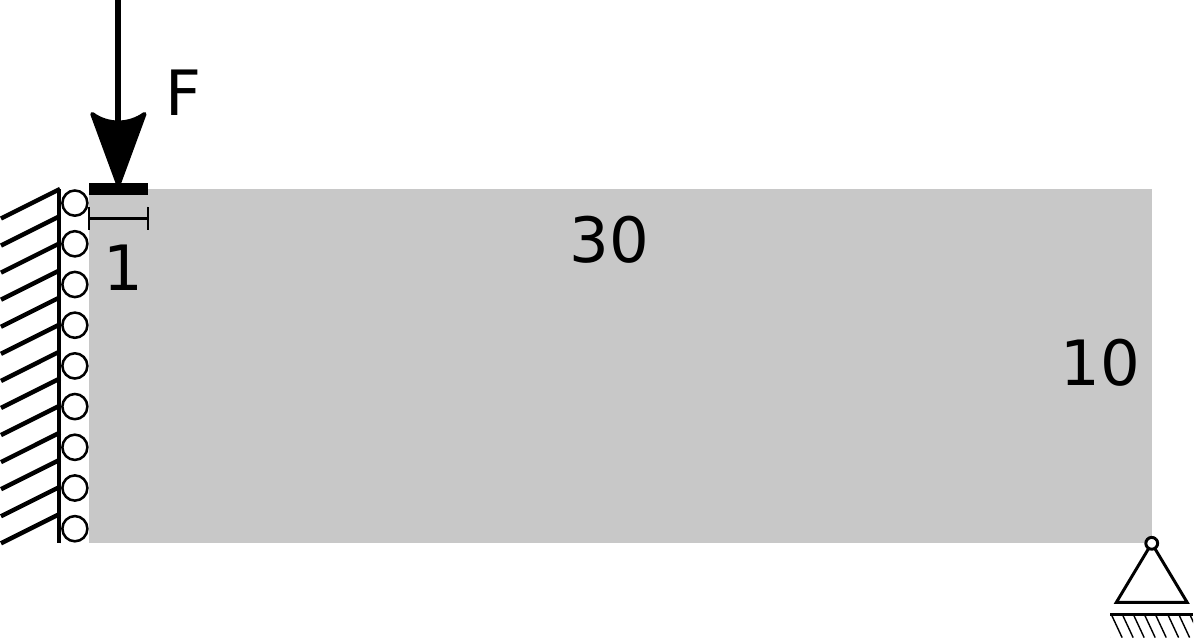}
		\caption{MBB beam.}
		\label{fig:second_sub}
	\end{subfigure}
	\caption{Setup of the benchmark examples tested in this work.}
	\label{fig:setup}
\end{figure}

The lack of penalization of intermediate densities effectively enlarges the space of admissible designs, offering the possibility of superior designs in terms of the chosen performance objective. In what follows, we compare the common benchmark examples of the cantilever beam and the MBB beam (Fig. \ref{fig:setup}). 

The cantilever beam is fully fixed on the left-hand edge and a distributed load $F$ is applied in the middle of the right-hand edge. The MBB beam setup exploits its symmetry and only the right half is considered, i.e., a symmetry boundary condition is applied to the left-hand edge. Instead of a point load, we consider a distributed load $F$ on the upper edge in the middle of the entire beam. The bottom right corner is restricted in displacement along the vertical axis. 
The standard compliance minimization problem under a volume fraction constraint is given as

\begin{equation}
\begin{split}
\min_{\forall \rho}  \ &: \ \mathcal{F}_{\textup{c}}\left(\rho, \tensor u\right) = \IntSB \tensor u\left(\rho\right) \cdot \tensor{t}_0 \text{\:dA}, \\
\textup{s.t.}\
&: \ \mathcal{G}_{\textup{vol}}\left(\rho\right) = \frac{\IntB \rho\left(\tensor{X}\right) \text{\:dV}}{V_0} - \overline{V}_{\rm frac} \leq \ 0, \\
&: \ 0 \leq \ \rho_e \leq 1 \ \ \ e = 1,...,N_e,
\end{split}
\label{eq:compliance}
\end{equation}

where $\mathcal{F}_{\textup{c}}$ is the compliance functional, $\mathcal{G}_{\textup{vol}}$ the volume fraction functional and $\overline{V}_{\rm frac}$ the volume fraction limit. For the optimizer, we employ the generalized optimality criteria method (GOCM) as proposed by \cite{kim2021generalized}. We justify the choice of GOCM by its surprising simplicity and robustness. We tested GOCM for compliance minimization problems with single and multiple constraints and each time obtained sensible results with stable convergence. The boundary value problem (BVP) is solved in the linear elastic regime and with an isotropic elastic material with a Young's modulus of $E = 1$ and a Poisson's ratio of $\nu = 0.3$.

\begin{figure*}[h!]
	\begin{subfigure}[b]{69mm} 
		\centering
		\includegraphics[width=0.9\linewidth]{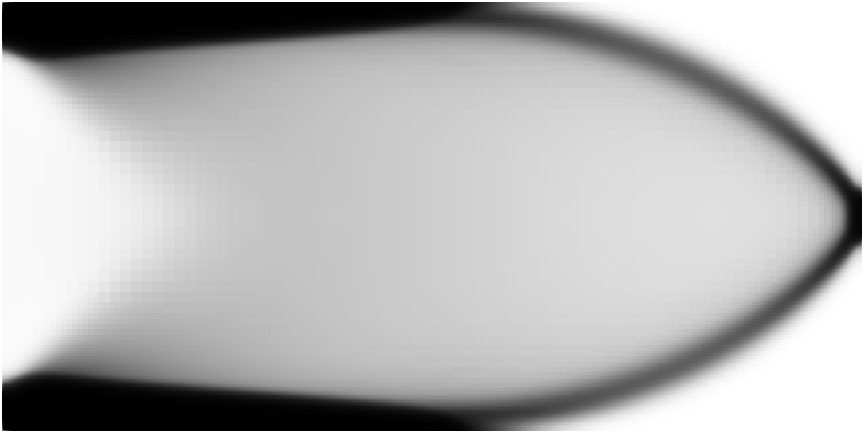}
		\caption{Cantilever; p = 1; $\mathcal{F}_{\textup{c}} = 7.61 \cdot 10^{-5}$.}
		\label{fig:c1}
	\end{subfigure}
	\hspace{2mm}
	\begin{subfigure}[b]{103mm} 
		\centering
		\includegraphics[width=0.9\linewidth]{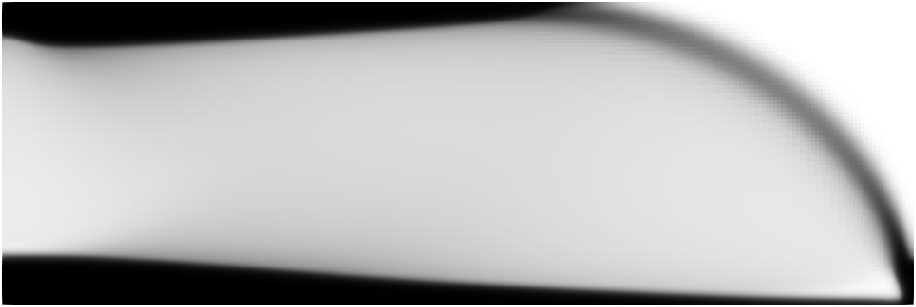}
		\caption{MBB; p = 1; $\mathcal{F}_{\textup{c}} = 2.16 \cdot 10^{-4}$.}
		\label{fig:m1}
	\end{subfigure}
	
	\begin{subfigure}[b]{69mm}
		\centering
		\includegraphics[width=0.9\linewidth]{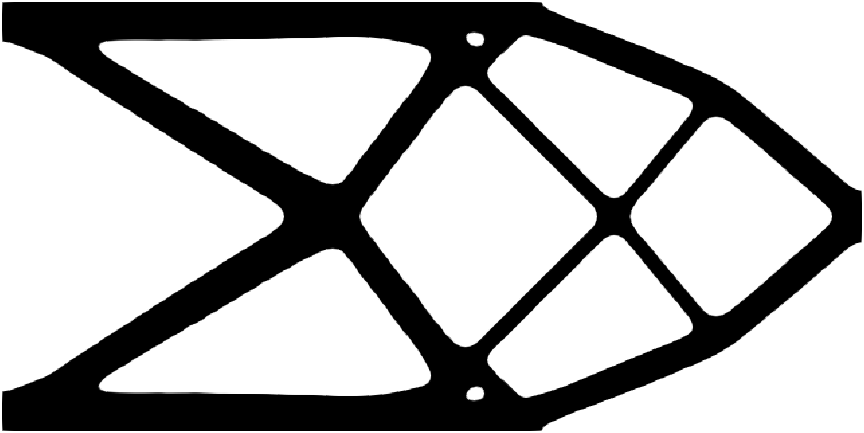}
		\caption{Cantilever; p = 3; $\mathcal{F}_{\textup{c}} = 8.92 \cdot 10^{-5}$.}
		\label{fig:c3}
	\end{subfigure}
	\hspace{2mm}
	\begin{subfigure}[b]{103mm} 
		\centering
		\includegraphics[width=0.9\linewidth]{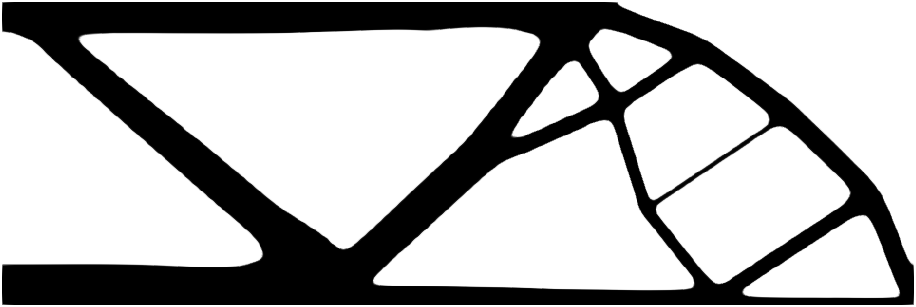}
		\caption{MBB; p = 3; $\mathcal{F}_{\textup{c}} = 2.65 \cdot 10^{-4}$.}
		\label{fig:m3}
	\end{subfigure}
	\caption{Comparison between the unpenalized (top) and penalized (bottom) topology optimization. The respective differences in the final compliance between the unpenalized and penalized cases are: $17.2\%$ for the cantilever and $22.7\%$ for the MBB beam.}
	\label{fig:penalization}
\end{figure*}

In Fig. \ref{fig:penalization}, we show the comparison between the unpenalized (variable thickness approach) and penalized topology optimization of the cantilever and MBB beam for a constrained volume fraction of $\overline{V}_{\rm frac} = 0.3$. As expected, penalizing the densities led to higher compliance values compared to the variable-thickness approach, that is, $17.2\%$ and $22.7\%$ for the cantilever and the MBB beam, respectively. 

A detailed study on sheet-like and truss-like structures was conducted in \cite{sigmund2016non}. For the 2D cantilever problem, unpenalized, penalized, and truss-based optimization methods were compared for compliance. For the 3D cantilever problem, a mesh and filter size study was conducted, revealing a tendency towards sheet-like structures as the element size decreases. In both 2D and 3D cases, truss-like structures exhibited compliance values approximately $30\%$ larger as compared to sheet-like structures. However, as noted in the same article, while sheet-like structures are favorable in terms of stiffness, additional constraints and requirements often force the optimizer to favor truss-like structures. Therefore, it is not valid to conclude that one form is ultimately superior to the other; rather, it boils down to the specific design goals and restrictions. With this in mind, our method aims to generate structures that bridge both worlds, addressing the compromised aspects of both types of structures, such as objective performance, manufacturability, and aesthetics.

\section{Multi-thickness method} \label{sec:method}

As in the previous section, we are interested in the design of structures in a 2D computational setting, where the assumed "extrusion" thickness is relatively small. However, as with the variable-thickness method, the structures should not be limited to the actual extrusion of 2D designs and should permit variations across the structure's thickness. 

Consider $n$ discrete target (allowable) thicknesses $t_1 < ... < t_i < ... < t_n$. For simplicity, we assume that $t_i = i t_1$; i.e., each thickness $t_i$ is a multiple of the first (and smallest) target thickness $t_1$. For this specific case, $\Delta t = t_1 = t_{i+1} - t_i$ for any $i \in [0,n-1]$ is the thickness interval, where $t_0 = 0$. In the context of density-based TO, we define the maximum thickness $t_n$ to correspond to the maximum target density $\hat{\rho}_n = 1$ and, for completeness, the void phase is noted as $\hat\rho_0 = 0$. Then the target densities that correspond to each target thickness are set to:

\begin{equation}
\hat\rho_i = \frac{t_i}{t_n} \hat\rho_n = \frac{t_i}{t_n}
\end{equation}

\begin{figure*}[h!]
	\centering
	\includegraphics[width=174mm]{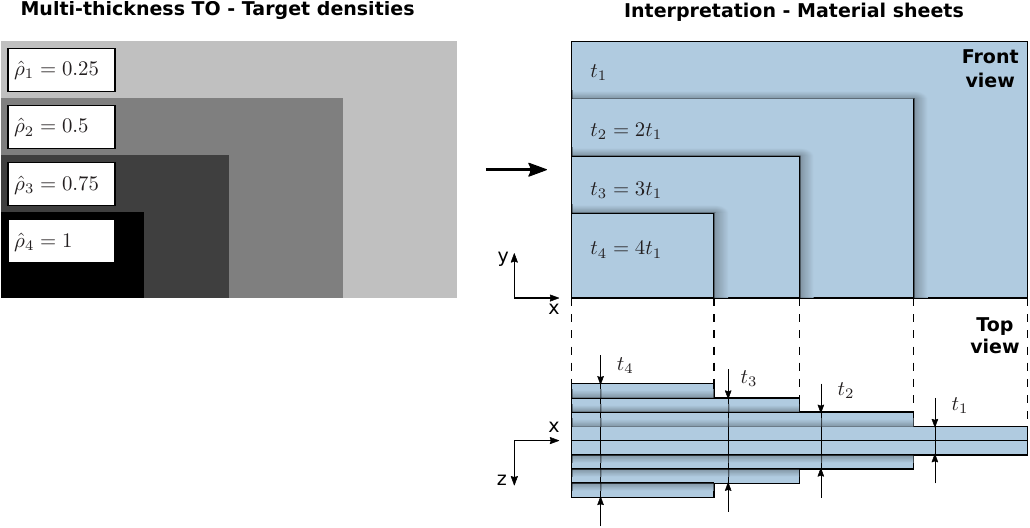}
	\caption{The concept and physical interpretation of the multi-thickness TO.}
	\label{fig:multithickness-concept}
\end{figure*}

With the multi-thickness method proposed here, one can consider a stack of independently cut sheets of material, each of constant thickness $t_1$. The visual interpretation of this concept is shown in Fig. \ref{fig:multithickness-concept}. Here, the result of the multi-thickness TO consists of only target densities $\hat\rho_i$, which correspond to their respective target thicknesses $t_i$. The interpreted structure is assembled by stacking together material sheets corresponding to the obtained shapes. In practice, symmetry with respect to the XY-plane needs to be fulfilled. Otherwise, unwanted behavior, such as out-of-plane deformation and twisting, would occur as a result of in-plane loading. Therefore, for each target density, we require two material sheets of thickness $0.5t_1$, which will be stacked on each side of the structure, i.e., symmetrically with respect to the XY-plane. 

Note that original density-based TO in 2D is equivalent to a single-thickness scenario with $t_1 = t_n$ and $\hat\rho_1 = 1$.

\subsection{Multilevel SIMP}

We propose a (local) $multilevel$ penalization technique to force the design variables toward the predefined target densities $\hat{\rho}_i$. The presence of density penalization naturally raises the point that the method is no longer of the "unpenalized" type. However, in the following sections, it should become clear that it is rather a combination of unpenalized and penalized versions of TO, both in terms of how the penalization is integrated in the design evolution and the impact on the design space and performance objectives. In short, this method can be thought of as \textit{bridging the gap} between the variable-thickness sheet and the standard SIMP approaches. 

We first recall the original (modified) SIMP formula from Eq. \eqref{eq:simp}. The multi-thickness formulation requires a generalization of the SIMP formula such that the densities are pushed toward the target densities $\hat\rho_i$, instead of only 0 (void) and 1 (solid). For any $\rho_e$ we determine the interval $\left[\hat\rho_i, \hat\rho_{i+1}\right]$ such that $\rho_e \in \left[\hat\rho_i, \hat\rho_{i+1}\right]$, then

\begin{equation}
E(\rho_e)=\left[\left[\frac{\rho_e - \hat\rho_i}{\hat\rho_{i+1} - \hat\rho_i}\right]^p\left[\hat\rho_{i+1} - \hat\rho_i\right] + \hat\rho_i\right] E_{0}
\end{equation}

where the penalized term represents the \textit{local} density

\begin{equation}
\rho_e^{\rm local} = \frac{\rho_e - \hat\rho_i}{\hat\rho_{i+1} - \hat\rho_i} \quad \text{for} \quad \rho_e \in \left[\hat\rho_i, \hat\rho_{i+1}\right].
\end{equation}

which gives

\begin{equation}
E(\rho_e)=\left[\left[\rho_e^{\rm local}\right]^p\left[\hat\rho_{i+1} - \hat\rho_i\right] + \hat\rho_i\right] E_{0}
\end{equation}

\begin{figure}[h!]
	\centering
	\includegraphics[width=84mm]{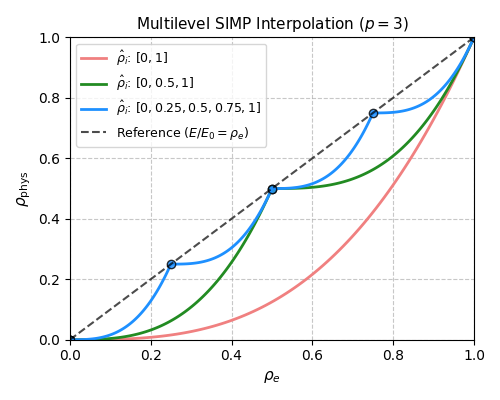}
	\caption{Multilevel SIMP plotted for $n = \{1,2,4\}$ number of target densities. Note that $n = 1$ is the standard SIMP rule.}
	\label{fig:multilevelsimp}
\end{figure}

The multilevel formulation is visualized in Fig. \ref{fig:multilevelsimp} for three sets of target densities, where the set $\hat\rho_i = \{0,1\}$ corresponds to the standard black-and-white SIMP model.

\begin{remark}[Continuity of multilevel SIMP]
	The multilevel formulation is locally continuous for each interval, with discontinuities at the respective target densities $\hat\rho_i$. Hence, the sensitivities do not exist at $\hat\rho_i$. Although this introduces a mathematical inconsistency into the proposed formulation, in a numerical setting, the problem is circumvented by applying an appropriate if-else statement to determine the active interval. The authors compared the proposed multilevel SIMP formulation with a more involved, smooth version of the multilevel SIMP, which is continuous at the target densities $\hat\rho_i$. However, the method proved to be less robust, since the target densities were not sufficiently fulfilled.
\end{remark}

\subsection{Multilevel smoothed Heaviside projection}

Regularization of the densities is a crucial step in topology optimization to circumvent its ill-posedness, which originates from mesh dependency leading to infeasible solutions, such as the checkerboard pattern. The most widely used strategy involves density filtering (here we use a PDE-based filter \cite{lazarov2011filters}) followed by projection using the smoothed Heaviside function \cite{wang2011projection}. The smoothed Heaviside projection sharpens the transition regions between solid and void, which are attenuated by the filtering step, using the following formula

\begin{equation}
H(\rho,\beta,\eta) = \frac{\tanh\left(\beta \eta\right) + \tanh\left(\beta \left[ \rho - \eta \right] \right)}{\tanh\left( \beta \eta \right) + \tanh \left( \beta \left[ 1 - \eta \right] \right)}
\label{eq:heaviside}
\end{equation}

where $\beta$ is responsible for the sharpness of the projection and $\eta$ is the projection threshold. Densities below the projection threshold $\eta$ are pushed towards 0 and those above $\eta$ are pushed towards 1. Naturally, for variable-thickness topology optimization, the projection is not applied, since the presence of intermediate densities is desired. However, for the multi-thickness approach, the projection can be beneficial if appropriately adapted to sharpen the transitions between target densities.

\begin{figure}[h!]
	\centering
	\includegraphics[width=84mm]{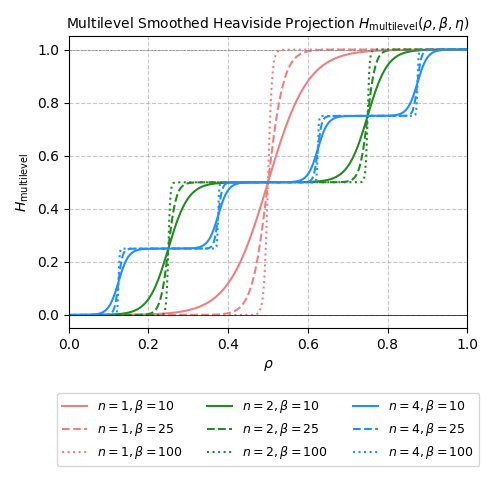}
	\caption{Multilevel smoothed Heaviside projection plotted for $n = \{1,2,4\}$ number of target densities and $\beta = \{10, 25, 100\}$ sharpness parameters. Note that $\beta_n = \beta \cdot n$.}
	\label{fig:multilevelheaviside}
\end{figure}

Thus, we define a multilevel smoothed Heaviside projection as follows

\begin{equation}
H_{\rm multilevel}(\rho,\beta,\eta) = \frac{1}{n} \sum_{i = 1}^n H_i^n (\rho, \beta_n, \eta_i^n)
\end{equation}

where the single projection functions

\begin{equation}
H_i^n(\rho,\beta_n,\eta_i) = \frac{\tanh\left(\beta_n \eta_i\right) + \tanh\left(\beta_n \left[ \rho - \eta_i \right] \right)}{\tanh\left( \beta_n \eta_i \right) + \tanh \left( \beta_n \left[ 1 - \eta_i \right] \right)}.
\end{equation}

are analogous to Eq. \eqref{eq:heaviside}, but incorporate the adjusted coefficients $\beta_n = \beta \cdot n$ and $\eta_i^n = \frac{i -0.5}{n}$ to retain the original sharpness of the projection and localize the threshold midway between the target densities. In Fig. \ref{fig:multilevelheaviside} the multilevel smoothed Heaviside projection is plotted for various numbers of target densities and various sharpness parameters $\beta$.

\subsection{Continuation strategy}\label{sec:continuation}

Multilevel penalization and multilevel Heaviside projection force the densities towards the target densities. Naturally, once the densities approach the nearest target density, it is difficult for them to "jump" to the next target density. Hence, to guarantee complete freedom in the first stage of optimization, the penalty is set to $p = 1$ and the projection sharpness to $\beta = 0.1$. This is equivalent to variable-thickness topology optimization. Once an intermediate stopping criterion is met, that is, after the densities have been free to converge towards any value in the range $\rho_e \in [0,1]$, the penalty continuation is activated, which effectively pushes the densities towards their nearest target densities $\hat\rho_i$. Once the continuation reaches the target penalty value $p = p_{max} = 3$, the continuation of the projection sharpness $\beta$ is initiated to improve the discreteness of the target thicknesses. The continuation strategy is defined as follows:

\begin{enumerate}
	\item Start with $p = p_{\rm init} = 1$ and $\beta = \beta_{\rm init} = 0.1$. 
	\item Once $\Delta \rho_{\rm mean} = \frac{1}{n_e}\sum_e \Delta \rho_e < 10^{-3}$ is fulfilled, start:
	\begin{equation}
	p^{I+1} = c_p p^I
	\end{equation}
	and continue until $p = p_{\rm max}$.
	\item Once $p = p_{\rm max}$ is fulfilled, start:
	\begin{equation}
	\beta^{I+1} = c_\beta \beta^I
	\end{equation}
	and continue until $\beta = \beta_{\rm max}$.
\end{enumerate}

where $I$ is the optimization iteration. The criterion $\Delta \rho_{\rm mean}$ is the mean density change and is also used as a stopping criterion for optimization, however, with the limit value of $10^{-4}$. In all numerical tests, we use the following parameters: $c_p = 1.03$, $c_\beta = 1.2$, $p_{\rm max} = 3$ and $\beta_{\rm max} = 50$. In the literature, projection sharpness values are often used up to $\beta_{\rm max} = 500$. However, paired with the filter radii of our choice, the value of $\beta_{\rm max} = 50$ already results in sufficiently sharp and well-defined transition regions between the respective target densities.

\subsection{Mesh adaptivity}\label{sec:meshadaptivity}

Mesh adaptivity is an effective way to boost computational efficiency, numerical accuracy and, in the context of TO, improve geometrical resolution. Hence, we employ an adaptivity technique based on a structured mesh as discussed in \cite{stankiewicz2025configurationalforcedrivenadaptiverefinementcoarsening} for TO and in \cite{stankiewicz2022geometrically, stankiewicz2024towards} for shape optimization using embedding domain discretization. However, we introduce new criteria for refinement and coarsening that are better suited for simple compliance problems within a linear elastic regime, as opposed to nonlinear, stress-constrained problems in \cite{stankiewicz2025configurationalforcedrivenadaptiverefinementcoarsening}. Thus, we perform h-adaptive refinement and coarsening on the basis of the density jump, given by:

\begin{equation}
\begin{split}
\text{Refinement:}\quad \left\llbracket \rho_e \right\rrbracket_{\max} &\geq c_r \Delta\hat\rho \\
\text{Coarsening:}\quad \left\llbracket \rho_e \right\rrbracket_{\max} &\leq c_c \Delta\hat\rho ,
\end{split}
\label{eq:ref}
\end{equation}

where $\left\llbracket \rho_e \right\rrbracket_{max}$ is the maximum density jump for the element $e$. That is, at each edge of the element $e$, the absolute density difference is calculated with the neighboring element that shares this edge. Of all the density differences between the element $e$ and its neighbors, the maximum value is chosen. The refinement and coarsening thresholds are defined by the difference between the target densities $\Delta\hat\rho$, multiplied by the parameters $c_r$, $c_c$. In this work, we use $c_r = 0.2$ and $c_c = 10^{-3}$ in all the numerical examples. The mesh adaptation takes place every fifth iteration of optimization. The feasibility of the adapted mesh is ensured by using the deal.II functionality for adaptive meshing. For details, refer to \cite{bangerth2007deal, arndt2021deal}.

\section{Numerical tests}\label{sec:tests}

In the following section, we test the multi-thickness approach using two common benchmark examples: the cantilever and the MBB beam. The setup of these problems is shown in Fig. \ref{fig:setup}. For each of the benchmarks, a varying number of target thicknesses and volume fractions are compared using the strategy shown in Table \ref{tab:cases}.

\begin{table}[h!]
	\centering
	\captionsetup{width=\columnwidth}
	\caption{Case study strategy for the cantilever and MBB beam. Each column corresponds to a common volume fraction, whereas each row corresponds to a common number of target thicknesses. The last row corresponds to the standard variable-thickness approach without penalization.}
	\label{tab:cases}
	\begin{tabular}{|c|c|c|}
		\hline
		\multicolumn{3}{|c|}{\textbf{$\bar V_{\rm frac}$ / $n_t$}} \\
		\hline
		0.2 / 1 & 0.3 / 1 & 0.5 / 1 \\
		0.2 / 2 & 0.3 / 2 & 0.5 / 2 \\
		0.2 / 3 & 0.3 / 3 & 0.5 / 3 \\
		0.2 / 4 & 0.3 / 4 & 0.5 / 4 \\
		0.2 / 8 & 0.3 / 8 & 0.5 / 8 \\
		0.2 / free & 0.3 / free & 0.5 / free \\
		\hline
	\end{tabular}
\end{table}

\begin{figure*}[h!]
	\centering
	\includegraphics[width=174mm]{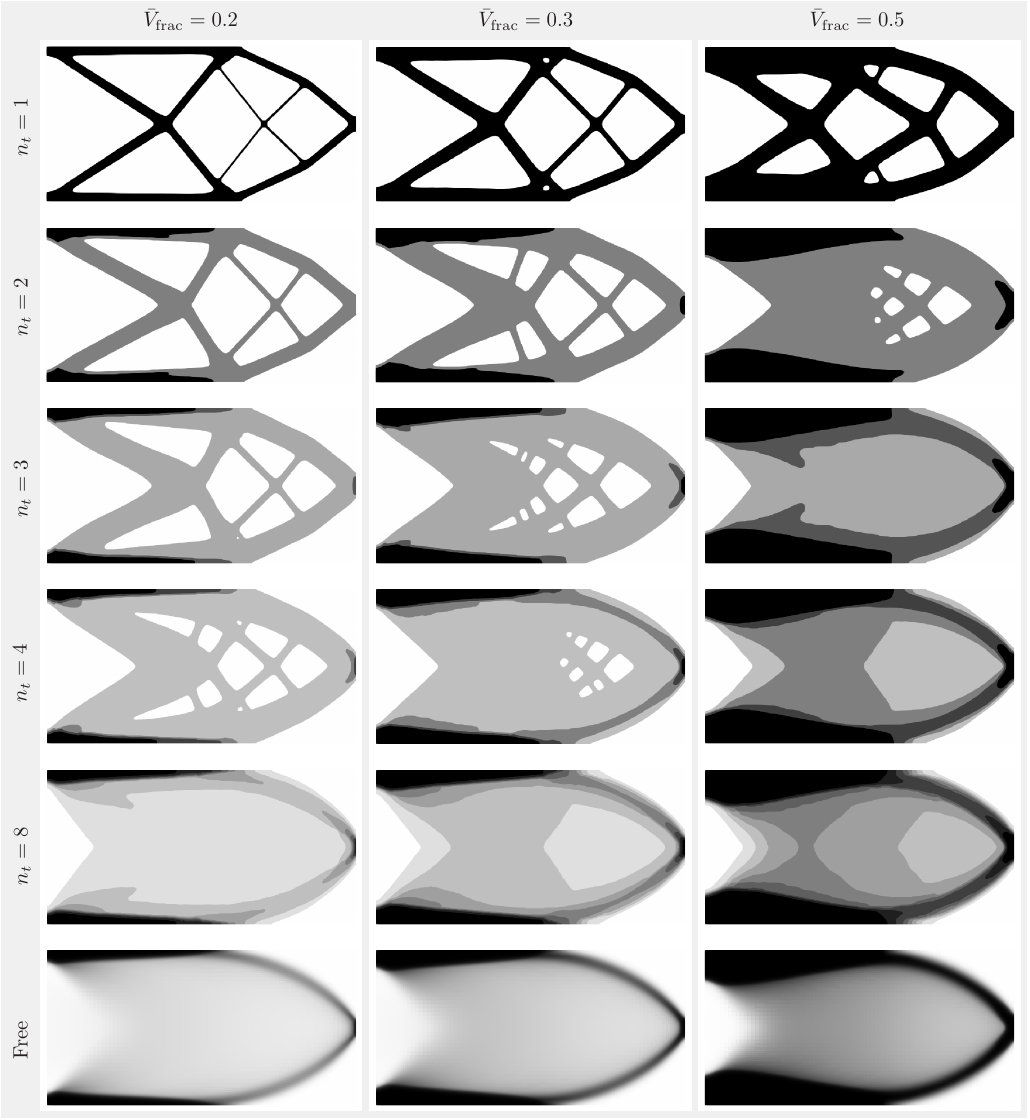}
	\caption{Case study for the cantilever benchmark according to the arrangement from Table \ref{tab:cases}. The single-target-thickness case ($n_t = 1$, the first row) is equivalent to standard TO. The last row shows the other extreme case which is the unpenalized, variable-thickness TO. The rows in between contain the multi-thickness cases ($n_t = 2 \to 8$). As the number of target thicknesses increases, the designs gradually transition from standard, penalized TO to unpenalized, variable-thickness TO.}
	\label{fig:cantilever-study}
\end{figure*}

\begin{figure*}[h!]
	\centering
	\includegraphics[width=174mm]{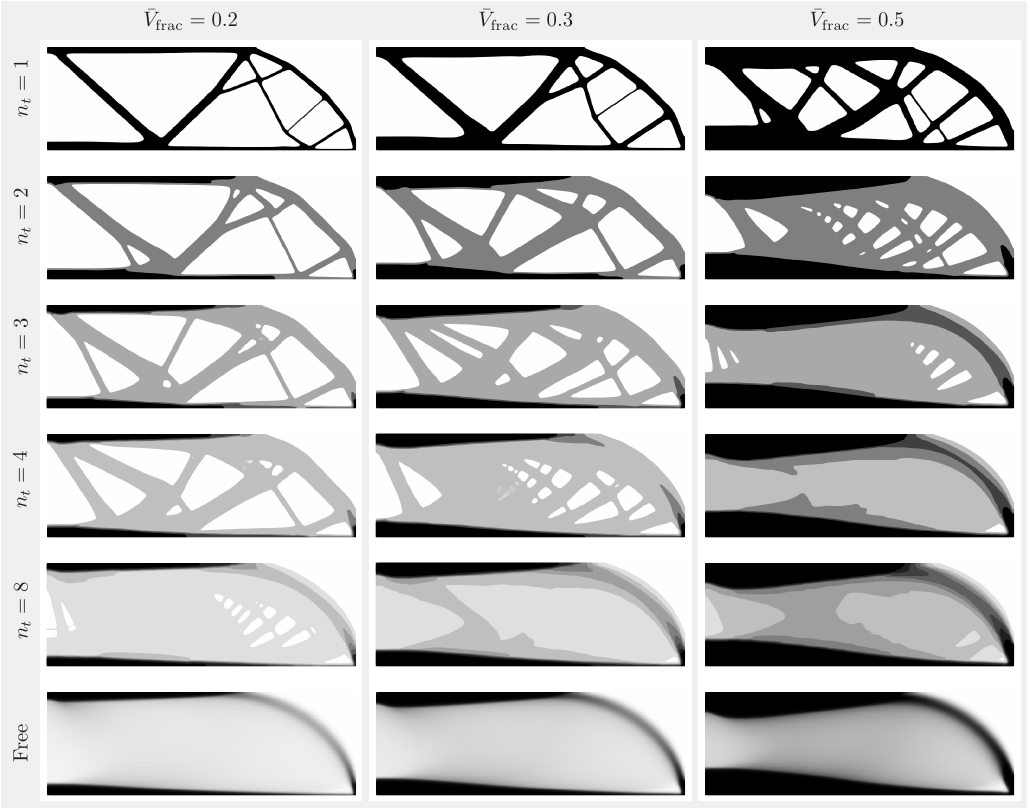}
	\caption{Case study for the MBB beam benchmark according to the arrangement from Table \ref{tab:cases}. The single-target-thickness case ($n_t = 1$, the first row) is equivalent to standard TO. The last row shows the other extreme case which is the unpenalized, variable-thickness TO. The rows in between contain the multi-thickness cases ($n_t = 2 \to 8$). As the number of target thicknesses increases, the designs gradually transition from standard, penalized TO to unpenalized, variable-thickness TO.}
	\label{fig:mbb-study}
\end{figure*}

\begin{figure}[h!]
	\centering
	\includegraphics[width=84mm]{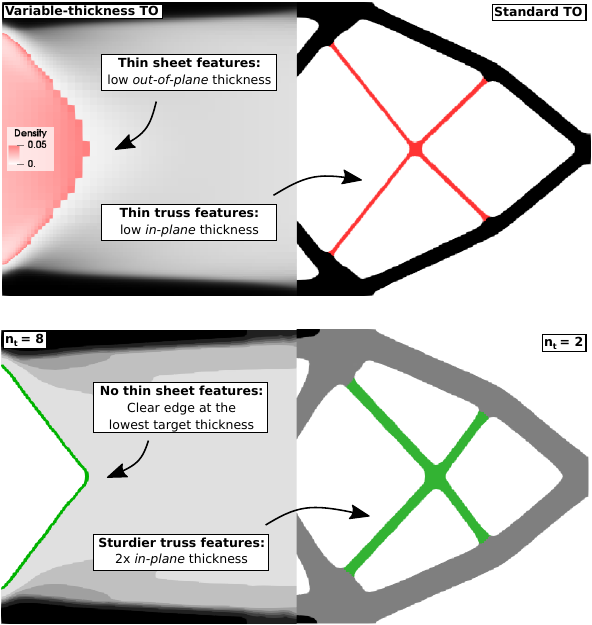}
	\caption{Thin sheet-like features are inherent to the variable-thickness TO, whereas thin truss-like features appear often in standard TO, in particular, when low volume fraction constraint is used ($\bar V_{\rm frac} \leq 0.3$). The multi-thickness approach completely eliminates thin sheet-like features and propotionally thickens truss-like features.}
	\label{fig:thin-features}
\end{figure}

\begin{figure*}[h!]
	\centering
	\includegraphics[width=174mm]{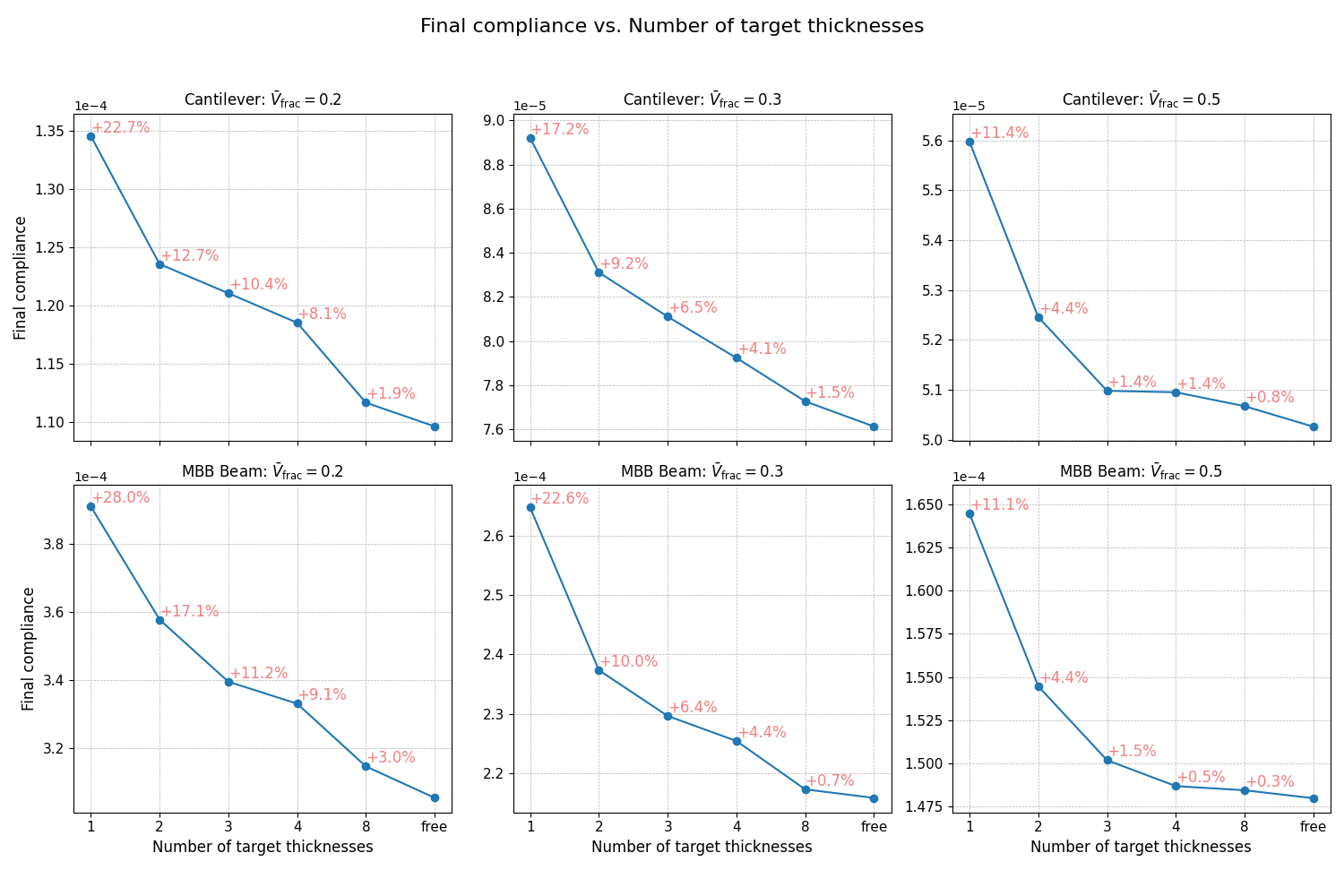}
	\caption{Tendency of the final compliance value with increasing number of target thicknesses. The red percentages inform about the difference between the current compliance value and the variable-thickness TO. The plots show a monotonically decreasing compliance value with an increasing number of target thicknesses. Already with a number of target thicknesses as small as $n_t = 3$, we obtain only $< 2 \%$ greater compliance as compared to the variable-thickness TO.}
	\label{fig:compliance-vs-n}
\end{figure*}

Note that the cases with $n_t = 1$ are equivalent to standard topology optimization, penalized using SIMP. The only exception is the p-continuation strategy in which the optimization starts with $p = 1$ instead of $p = 3$. The material properties, the optimizer, and the definition of the optimization problem are the same as described in Section \ref{sec:topopt}. The mesh adaptation follows the strategy described in Section \ref{sec:meshadaptivity}. The initially generated meshes are 20x10 for the cantilever and 30x10 for the MBB beam, respectively, and are then uniformly refined twice, providing 80x40 and 120x40 starting meshes. In other words, we start with meshes that have a uniform refinement level of 2. We set the maximum refinement level to 5 and the minimum refinement level to 0. The filter radius for the PDE filter equals 0.375, that is 1.5 times the element length at refinement level 2. For the stopping criteria, we employ the mean design change, defined as:

\begin{equation}
\Delta\rho_{\rm mean} = \frac{\sum_{e=1}^{n_e} \Delta \rho_e}{n_e} \le \varepsilon
\end{equation}

and the maximum number of iterations $I \ge I_{\max}$, depending on which criterion is met first. We select $I_{\max} = 200$ and $\varepsilon = 10^{-4}$.

Finally, in Figs. \ref{fig:cantilever-study} and \ref{fig:mbb-study}, the results of the cantilever and MBB beam case studies are shown, respectively. 

First, we note that the cases with $n_t = 1$ behave exactly as standard penalized TO. As the number of target thicknesses increases, the designs gradually shift toward the non-Michell type of structure and become more similar to variable-thickness TO designs. Interestingly, topological changes (holes) stop occurring for a smaller number of target thicknesses as the volume fraction constraint increases. For example, for the cantilever cases with $\bar V_{\rm frac} = 0.5$, the designs with $n_t = 3$ and higher already consist of only continuous unperforated sheets. However, at the other end of the spectrum, for the MBB beam cases with $\bar V_{\rm frac} = 0.2$, Michell-type features appear even with $n_t = 8$. In terms of in-plane dimensions, very thin features are avoided for $n_t \geq 2$, which is a fundamental advantage in terms of manufacturability and durability. This is particularly pronounced for the designs with low volume fraction constraints ($\bar V_{\rm frac} \leq 0.3$). These thin features in the standard TO designs ($n_t = 1$) appear despite using a PDE filter with a filter radius significantly larger than the thickness of the resulting features, which is likely a consequence of a very fine mesh in these regions. In fig. \ref{fig:thin-features} we focus on the observation that the multi-thickness method successfully alleviates the presence of both thin sheet-like and truss-like features. The comparison between $n_t / \bar V_{\rm frac} = \rm free / 0.2$ and $n_t / \bar V_{\rm frac} = 8 / 0.2$ reveals the elimination of impractical, thin sheets, replaced by a clearly defined edge. Whereas the comparison between $n_t / \bar V_{\rm frac} = 1 / 0.2$ and $n_t / \bar V_{\rm frac} = 2 / 0.2$ shows how, considering the in-plane dimensions, the truss-like features are approximately two times thicker if the out-of-plane thickness decreases by the same proportion.

The asymmetry in the designs is an artifact of the mesh adaptivity process. Specifically, the algorithms that ensure a feasible node connectivity do not account for the structure's symmetry. Consequently, while cells may be marked symmetrically for refinement or coarsening, the final adapted mesh is not guaranteed to be symmetrical, leading to uneven evolution of the designs.

A crucial finding is that the designs obtained using the multi-thickness strategy bear a very close resemblance to the 3D cantilever designs obtained in \cite{sigmund2016non}, both in the mesh refinement and volume fraction study, that is, from Figs. 2 and 3 in \cite{sigmund2016non}. It is clear that as we increase the number of target thicknesses $n_t$ in the multi-thickness approach, the designs evolve similarly to those in the mesh refinement study using 3D TO, and both approaches show a similar design evolution with the increase of the volume fraction constraint. 

In \cite{sigmund2016non}, the 3D study demonstrated the general superiority of sheet-like structures compared to the penalized, and therefore compromised, Michell-like designs. In this work, it is clear that designs with $n_t > 1$ are at an intermediate stage between standard penalized TO and variable-thickness TO. Thus, in order to evaluate the performance of the designs obtained using the multi-thickness approach, the final objective values for each design case are plotted against the number of target thicknesses on the horizontal axis in Fig. \ref{fig:compliance-vs-n}.

\begin{figure}[h!]
	\begin{subfigure}[b]{84mm} 
		\centering
		\includegraphics[width=0.7\linewidth]{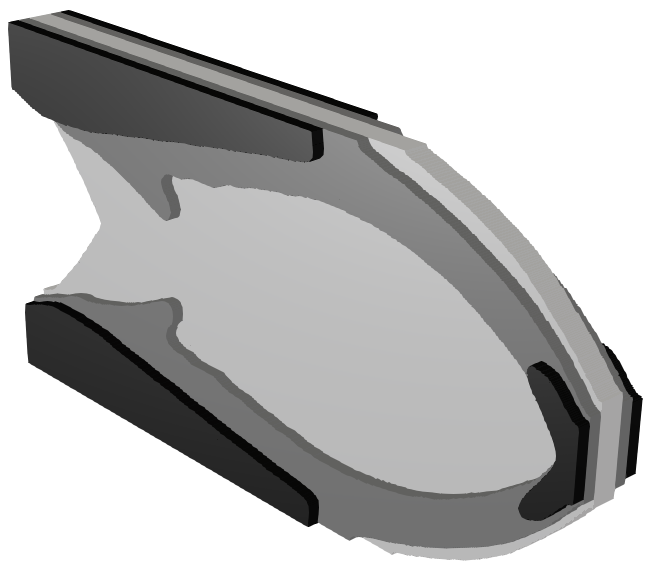}
		\caption{Perspective view of a joint model.}
		\label{fig:joint}
	\end{subfigure}
	\hspace{0.02\textwidth}
	\begin{subfigure}[b]{84mm} 
		\centering
		\includegraphics[width=\linewidth]{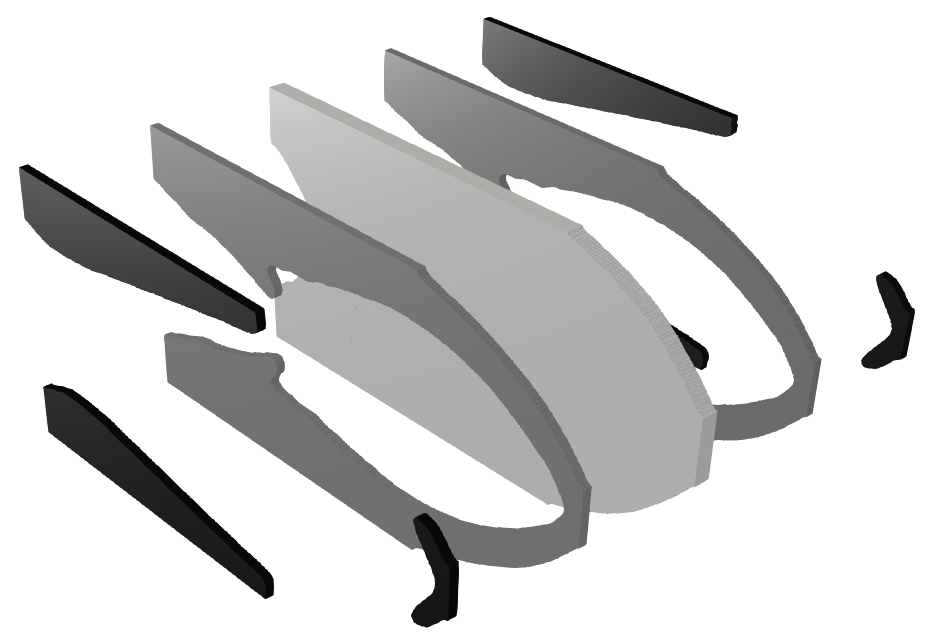}
		\caption{Assembly of single material sheets.}
		\label{fig:assembly}
	\end{subfigure}
	\caption{Thickness interpretation of the $n_t / \bar V_{\rm frac} = 3 / 0.5$ cantilever design. The coloring of each material sheet is consistent with the target densities. Note that the middle sheet is of double thickness of a single sheet.}
	\label{fig:3d-example}
\end{figure}

\begin{figure*}[h!]
	\begin{subfigure}[t]{84mm} 
		\centering
		\includegraphics[width=0.8\linewidth]{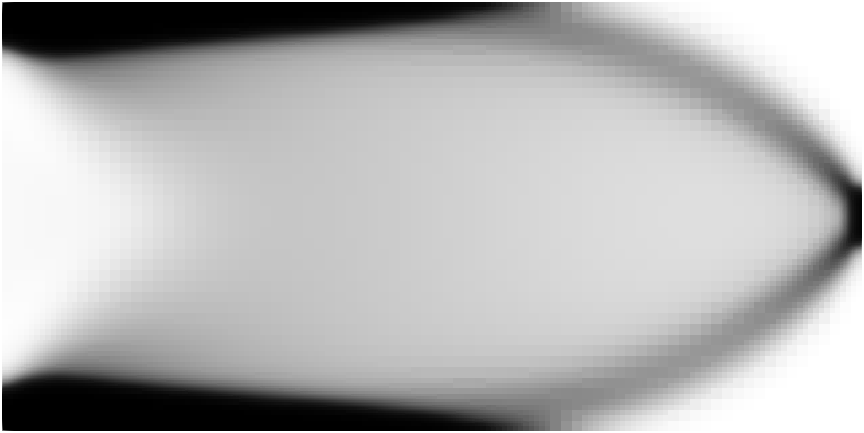}
		\caption{Iteration 27: Start of p-continuation: $p = 1$, $\beta = 0.1$.}
		\label{fig:c-evo-1}
	\end{subfigure}
	\begin{subfigure}[t]{84mm} 
		\centering
		\includegraphics[width=0.8\linewidth]{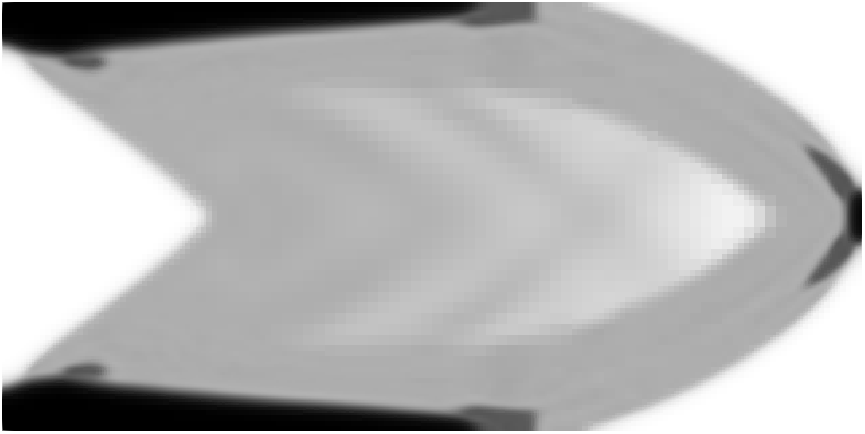}
		\caption{Iteration 65: End of p-continuation, start of $\beta$-continuation: $p = 3$, $\beta = 0.1$.}
		\label{fig:c-evo-2}
	\end{subfigure}
	
	\begin{subfigure}[b]{84mm}
		\centering
		\includegraphics[width=0.8\linewidth]{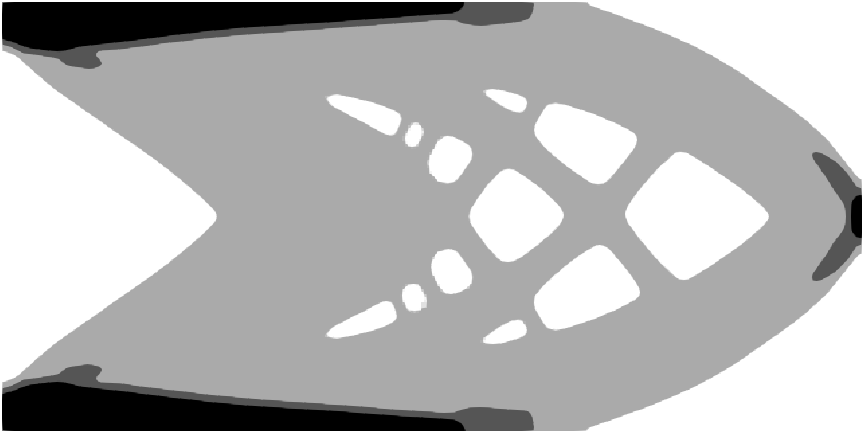}
		\caption{Iteration 100: End of $\beta$-continuation: $p = 3$, $\beta = 50$.}
		\label{fig:c-evo-3}
	\end{subfigure}
	\begin{subfigure}[b]{84mm} 
		\centering
		\includegraphics[width=0.8\linewidth]{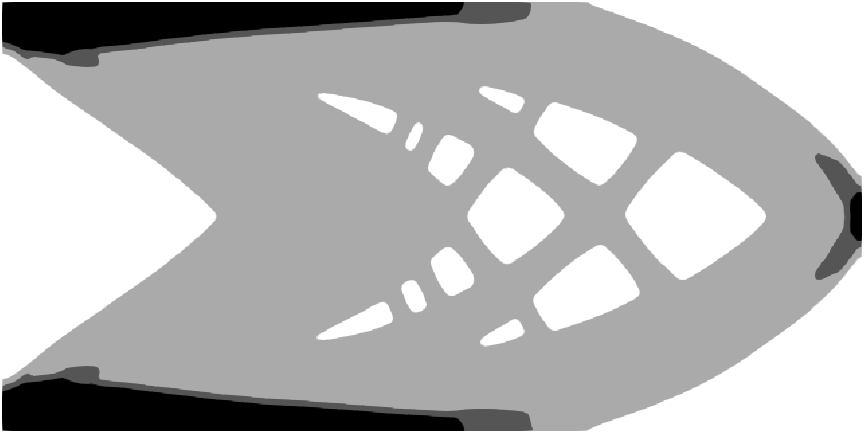}
		\caption{Iteration 200: Iteration limit reached.}
		\label{fig:c-evo-4}
	\end{subfigure}
	\caption{Evolution of the cantilever design with $n_t = 3$ and $\bar V_{\rm frac} = 0.3$. Up to iteration 27, the problem behaves like the known variable-thickness optimization. During p-continuation (iterations 27-65), the densities are forced towards the target values, yet the design is not clearly defined. During the $\beta$-continuation (iterations 65-100), the final topology and clear, sharp boundaries of the thickness layers form. In iterations 100-200, the final fine-tuning of the design takes place.}
	\label{fig:c-evo}
\end{figure*}

In addition, for each of the designs with a fixed number of target thicknesses, a percentage difference in the final objective value is calculated compared to the variable-thickness approach. The designs with $n_t = 1$, equivalent to standard penalized TO, naturally exhibit a higher final compliance compared to variable-thickness TO, with the difference being in the range of $11.1 - 28.0\%$, where the higher the volume fraction constraint, the smaller the difference. Interestingly, enabling $n_t = 2$ already reduces this difference to the range of $4.4 - 17.1\%$. The designs with $n_t = 3$ reduce it further to $~1.5\%$ in the case of $\bar V_{\rm frac} = 0.5$. In relation to this, as observed in Figs. \ref{fig:cantilever-study} and \ref{fig:mbb-study}, these designs already consist only of continuous unperforated sheets. Consequently, we can clearly see that the multi-thickness designs without Michell-type structures exhibit compliance values that fall very close to the variable-thickness cases. Moreover, based on the slope of the plots in Fig. \ref{fig:compliance-vs-n}, significant jumps in the final compliance difference occur between the designs with and without Michell-type features, for example, the cases $n_t / \bar V_{\rm frac} = 4 / 0.2$ vs. $n_t / \bar V_{\rm frac} = 8 / 0.2$ (cantilever) or $n_t / \bar V_{\rm frac} = 2 / 0.5$ vs. $n_t / \bar V_{\rm frac} = 3 / 0.5$ (cantilever and MBB beam). Generally, the final objective values of all the multi-thickness designs that consist of continuous unperforated sheets fall within $2\%$ of the final objective value of the variable-thickness approach.

In Fig. \ref{fig:3d-example}, we show an exemplary 3D model of the $n_t / \bar V_{\rm frac} = 3 / 0.5$ cantilever case. By choosing a total thickness of $t = 20\rm mm$ for a 200x100mm planar design, the single sheet thickness equals $t_i = 0.5t / n_t = 3.33\rm mm$. Note that the middle sheet in Fig. \ref{fig:3d-example} consists of two single sheets. This demonstrates that not only additive manufacturing but also more conventional techniques, such as sheet cutting and joining, can be used to produce such 3D structures.

Due to the continuation strategy, as described in Section \ref{sec:continuation}, the multi-thickness optimization consists of four stages: (i) unpenalized optimization, (ii) p-continuation, (iii) $\beta$-continuation, (iv) fine-tuning. This strategy ensures the largest design space and freedom of design changes before penalization toward target densities is introduced. In Fig. \ref{fig:c-evo}, the intermediate designs at the end of each of the four stages are shown. The first stage was already completed at iteration 27, as the intermediate stopping criterion $\Delta \rho_{\rm mean} < 10^{-3}$ was met. Until then, the method behaves like a variable thickness TO (Fig. \ref{fig:c-evo-1}). The p-continuation stage takes 38 iterations to reach $p = 3$. The densities were then pushed toward their nearest target values. A rough sheet-like structure is forming, yet without any fine details and clear boundaries of the respective sheets (Fig. \ref{fig:c-evo-2}). During the $\beta$-continuation stage, which takes 35 iterations, the design is clearly formed, including Michell-type features, as shown in Fig. \ref{fig:c-evo-3}. In the last stage, minor tweaks are made to the design until a stopping criterion is met.

\begin{figure}[h!]
	\begin{subfigure}[b]{82mm} 
		\centering
		\includegraphics[width=\linewidth]{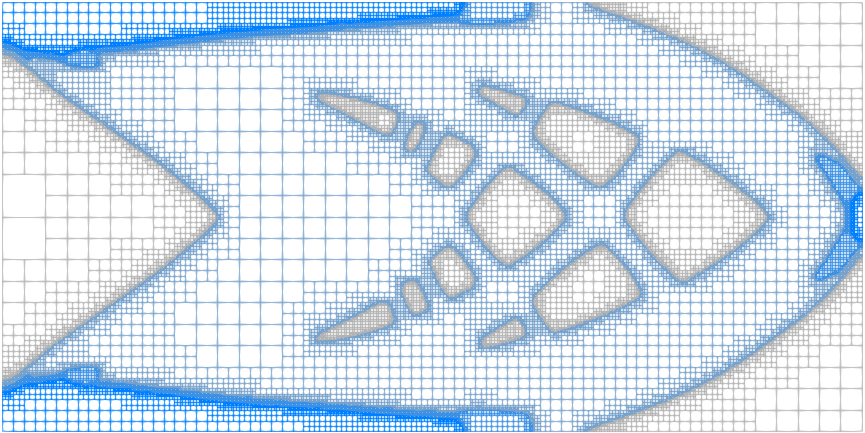}
		\caption{Final mesh for the case: $n_t/\bar V_{\rm frac} = 3/0.3$.}
		\label{fig:v3-mesh}
	\end{subfigure}
	\vspace{0.02\textwidth}
	\begin{subfigure}[b]{82mm} 
		\centering
		\includegraphics[width=\linewidth]{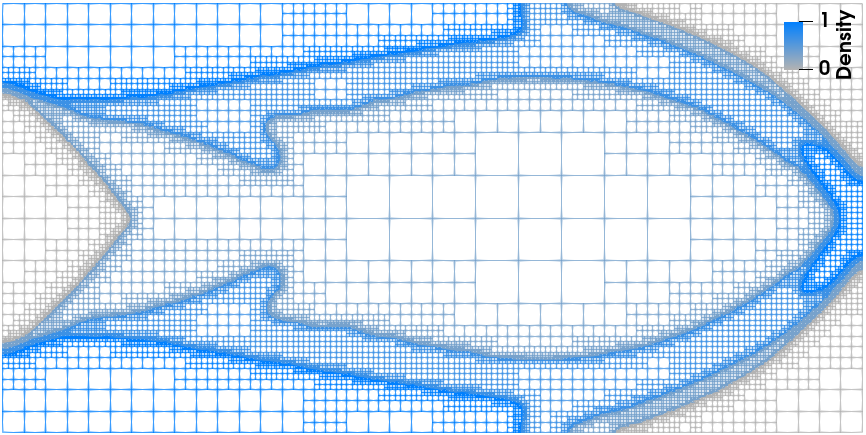}
		\caption{Final mesh for the case: $n_t/\bar V_{\rm frac} = 3/0.5$.}
		\label{fig:v5-mesh}
	\end{subfigure}
	\caption{Final meshes of two study cases using the adaptive mesh strategy from \ref{sec:meshadaptivity}. The model is displayed using a wireframe style and the densities are represented by a blue-to-grey scale to facilitate the visibility of the void regions. The final meshes have 28,145 and 23,732 cells, respectively.}
	\label{fig:cantilever-mesh}
\end{figure}

For performance considerations, it is relevant to assess the effectiveness of the mesh adaptivity strategy from Section \ref{sec:meshadaptivity}. Using an adaptive mesh in TO has a significant impact on computational cost and geometric accuracy of the designs \citep{stankiewicz2025configurationalforcedrivenadaptiverefinementcoarsening}. As mesh adaptivity is not the main focus of this work, our assessment here is limited to a visual inspection and the number of cells in the final designs. In Fig. \ref{fig:cantilever-mesh}, two selected cases are shown in a wireframe style with a blue-to-grey density scale. A brief visual inspection suggests that the employed mesh adaptivity strategy perfectly suits the multi-thickness approach. The mesh is consistently fine along the boundaries of the material sheets and coarse away from them. This facilitates very fine geometrical features and clear boundaries, while maximizing computational performance by maintaining a coarse mesh everywhere else. The final meshes shown in Figs. \ref{fig:v3-mesh} and \ref{fig:v5-mesh} contain 28,145 and 23,732 cells, respectively. In order to obtain the same level of detail, a uniform mesh would require a 640x320 grid, that is, 204,800 cells. 

\section{Manufacturing}\label{sec:am}

\begin{figure*}[h!]
	\begin{subfigure}[t]{55mm} 
		\centering
		\includegraphics[width=\linewidth]{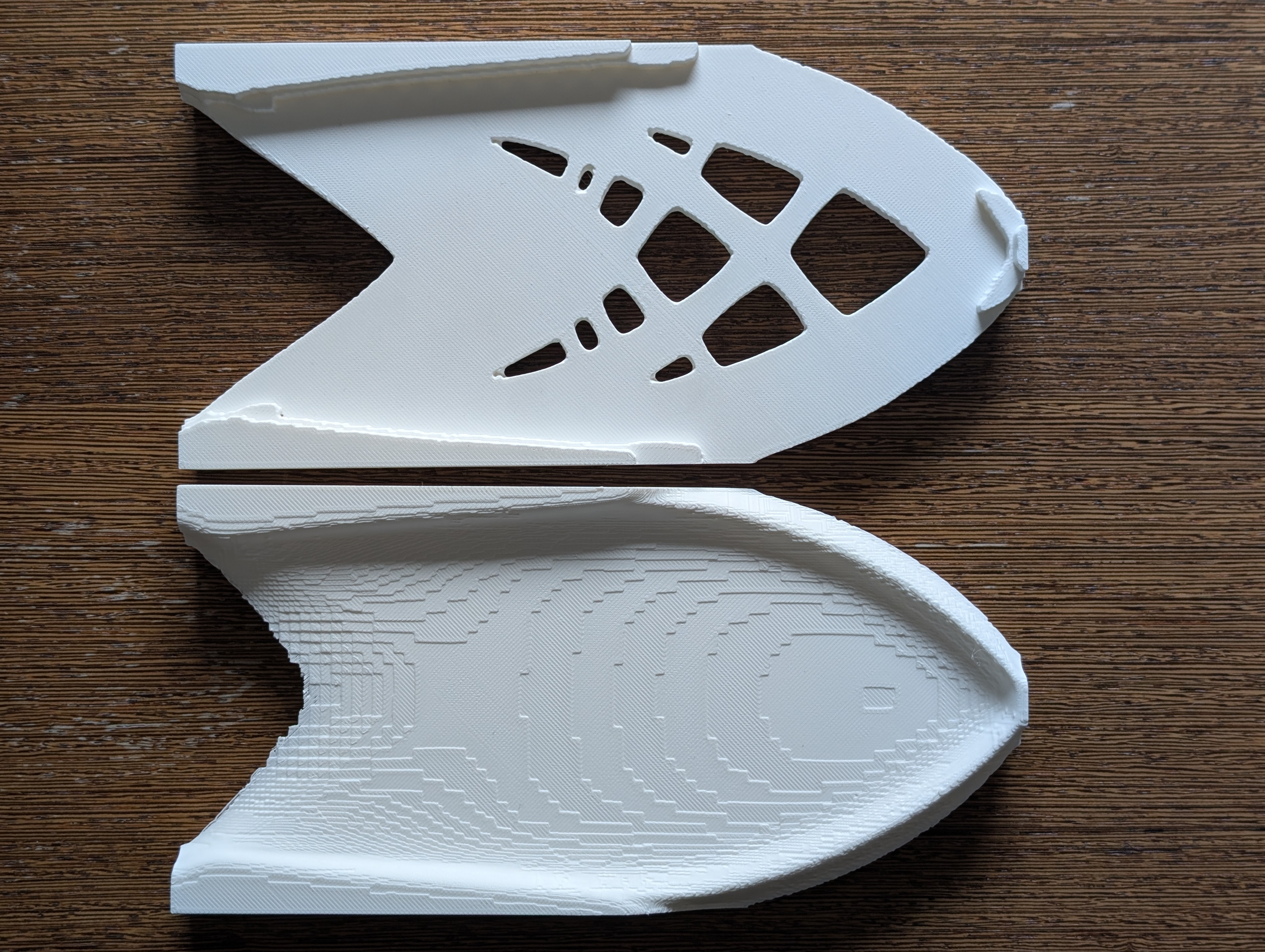}
		\caption{Top view of both structures.}
		\label{fig:print-both}
	\end{subfigure}
	\vspace{0.02\textwidth}
	\begin{subfigure}[t]{55mm} 
		\centering
		\includegraphics[width=\linewidth]{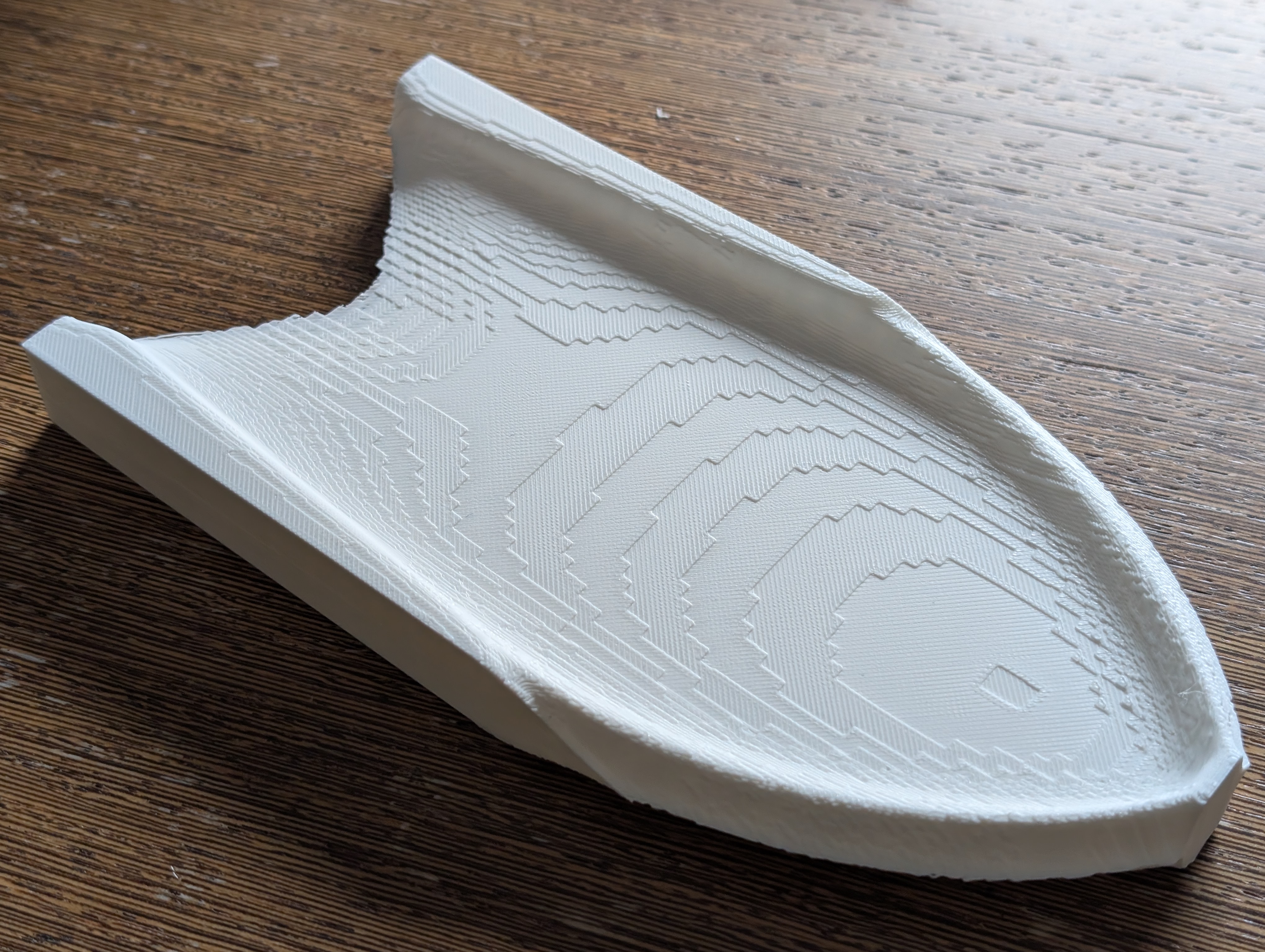}
		\caption{Printed variable-thickness structure: $n_t / \bar{V}_{\mathrm{frac}} = \mathrm{free}/0.3$.}
		\label{fig:print-vt}
	\end{subfigure}
	\vspace{0.02\textwidth}
	\begin{subfigure}[t]{55mm}
		\centering
		\includegraphics[width=\linewidth]{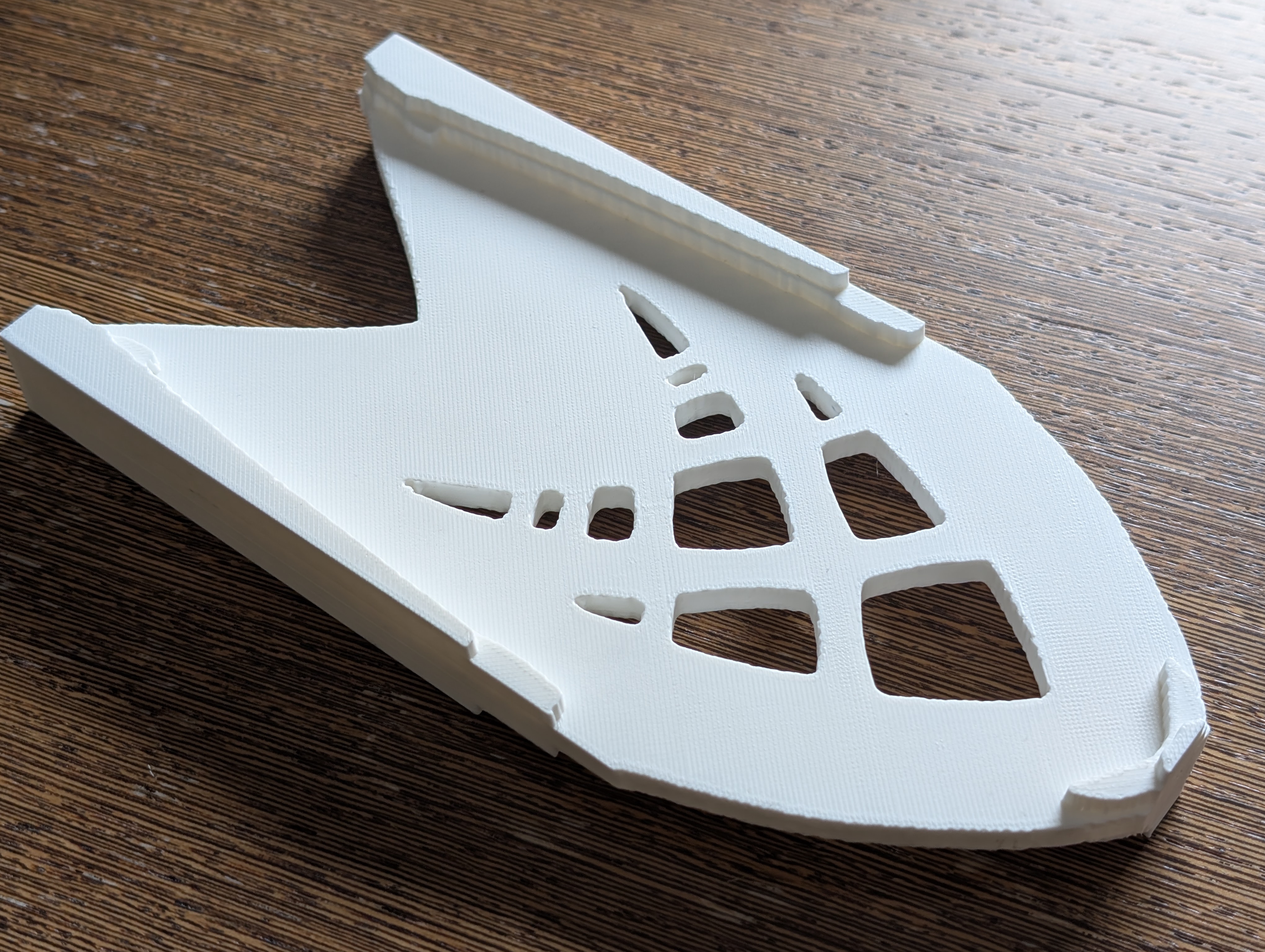}
		\caption{Printed multi-thickness structure: $n_t / \bar{V}_{\mathrm{frac}} = 3/0.3$.}
		\label{fig:print-mt}
	\end{subfigure}
	\caption{Additive manufacturing of variable-thickness ($n_t / \bar{V}_{\mathrm{frac}} = \mathrm{free}/0.3$) and multi-thickness ($n_t / \bar{V}_{\mathrm{frac}} = 3/0.3$) structures using Bambu Lab P1S 3D printer.}
	\label{fig:print}
\end{figure*}

To evaluate practical aspects of the multi-thickness approach, two cantilever designs were additively manufactured: one with three discrete thickness levels ($n_t / \bar{V}_{\mathrm{frac}} = 3/0.3$) and one with continuous thickness variation ($n_t / \bar{V}_{\mathrm{frac}} = \mathrm{free}/0.3$). Both models were printed using a Bambu Lab P1S 3D printer equipped with a 0.4mm nozzle, and standard PLA filament. As no mechanical testing was planned, the infill density was kept at the default value of 15\%, since increasing it to 100\% would not affect surface patterns but would increase print time by roughly 20\%. The focus was placed on surface quality and printing efficiency. A layer height of 0.16mm, half the recommended maximum, was used to ensure consistent surface resolution. The planar dimensions of the structures were scaled to 200mm x 100mm, and the maximum thickness was set to 19.2mm. For the $n_t = 3$ case, this resulted in a target thickness level of 3.2mm (corresponding to three full layers per side), which aligns with the selected layer height. The printed structures are shown in Fig. \ref{fig:print}.

The printing times for the variable-thickness and multi-thickness models were 275 and 228 minutes, respectively, indicating a 20\% longer print time for the variable-thickness structure, even though the multi-thickness model featured more complex topological details. \footnote{For a multi-thickness structure without topological features, such as $n_t / \bar V_{\rm frac} = 3/0.5$, the corresponding variable-thickness design with the same volume fraction ($n_t / \bar V_{\rm frac} = \rm free/0.5$) would take approximately 28\% longer to print.} 

Both structures required support material to accommodate overhanging features. However, the detachment process was significantly easier for the multi-thickness model due to the presence of flat surfaces. In contrast, the curved geometry of the variable-thickness structure resulted in visible staircase patterns on the surface, a common artifact when printing smooth curvature using layered deposition. This suggests that variable-thickness designs may require additional post-processing, such as surface smoothing or higher-resolution slicing, to achieve comparable surface quality. In the case of the multi-thickness structure, these artifacts did not occur, as each flat surface was completed within a single print layer, yielding a much cleaner finish.

\section{Conclusions and outlook}\label{sec:conclusion}

In this work, we have presented a novel multi-thickness method for density-based topology optimization to generate high-performance, manufacturable structures. By generalizing the SIMP penalization and Heaviside projection to operate across multiple, predefined thickness levels, our method successfully guides the optimization towards designs composed of discrete, physically meaningful sheet thicknesses.

The key findings from our numerical studies on cantilever and MBB beam problems are as follows:

\bmhead{Performance} The multi-thickness approach effectively bridges the performance gap between standard penalized TO and unpenalized variable-thickness optimization. We have shown that by introducing even a small number of discrete thicknesses ($n_t \ge 3$), the resulting designs achieve compliance values that are remarkably close (often differing by less than $2\%$) to the reference optimum of a fully variable-thickness sheet. This demonstrates that a significant portion of the performance benefits can be captured without requiring a continuous range of thicknesses.

\bmhead{Design Morphology} The number of target thicknesses directly influences the final structural topology. A single thickness ($n_t = 1$) replicates standard SIMP results, producing truss-like structures. As $n_t$ increases, the designs progressively incorporate more sheet-like features, eventually converging towards the morphology of unpenalized solutions. This provides the designer with a powerful parameter to control the trade-off between traditional truss-like aesthetics and optimal sheet-based forms.

\bmhead{Manufacturability} The method inherently enhances manufacturability. By eliminating regions of near-zero thickness in the out-of-plane direction, characteristic to variable-thickness designs, and restricting the design to a set of specified thicknesses, it produces structures that are more robust and less prone to buckling. Furthermore, the Michell-type features in the multi-thickness structures are inherently thicker within the in-plane directions, offering another practical advantage from the manufacturing and buckling point of view. Additionally, the resulting designs are well-suited for fabrication via the assembly of individually cut standard-thickness sheets or through more efficient additive manufacturing processes where layering strategies can be simplified. Furthermore, additive manufacturing of variable-thickness and multi-thickness structures demonstrated clear advantages of the multi-thickness approach in terms of reduced print time, improved surface quality, and easier detachment of support structures. 
\\

The integration of an adaptive meshing strategy proved highly effective, enabling fine resolution of the boundaries between different thickness regions while maintaining computational efficiency. In summary, the proposed multi-thickness optimization framework is a versatile tool that offers a practical path to designing structures that do not compromise performance for the sake of manufacturability.

Future work will focus on implementing advanced physics and responses for the multi-thickness method. In particular, the way stresses are calculated within variable-thickness TO differs from standard TO. Therefore, a realistic formulation of stresses within that context is required. Moreover, robust post-processing techniques are necessary, which would offer various possibilities to generate sidewalls of the thickness sheets, applicable for additive manufacturing. For instance, controlling the angle at which the sidewalls join adjacent material sheets is crucial to prevent stress concentrations.

\bmhead{Acknowledgements}
Funded by the European Union. Views and opinions expressed are however those of the author(s) only and do not necessarily reflect those of the European Union or the European Research Council Executive Agency. Neither the European Union nor the granting authority can be held responsible for them. This work is supported by the European Research Council (ERC) under the Horizon Europe program (Grant-No. 101052785, project: SoftFrac).
\begin{figure}[h!]
	\centering
	\includegraphics[width=0.3\textwidth]{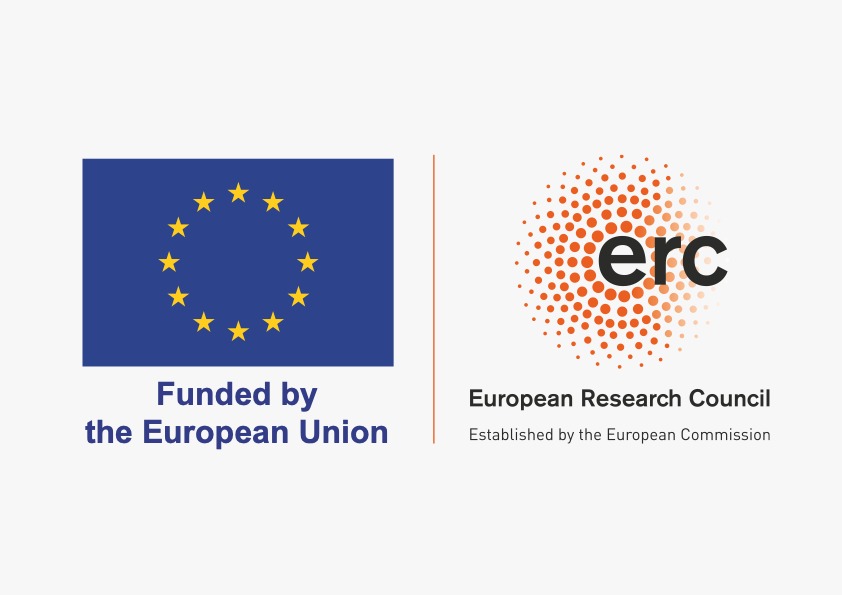}
\end{figure}

The authors gratefully acknowledge the assistance of Jan Hinrichsen with the 3D printing of the structures used in this study.
\section*{Declarations}
\bmhead{Author contributions}
Conceptualization: Gabriel Stankiewicz; Methodology: Gabriel Stankiewicz; Implementation: Gabriel Stankiewicz and Chaitanya Dev; Discussions and improvement of the methodology: all authors; Manufacturing: Gabriel Stankiewicz; Writing - original draft preparation: Gabriel Stankiewicz; Writing - review and editing: all authors; Funding acquisition: Paul Steinmann, Resources: Paul Steinmann, Supervision: Paul Steinmann.
\bmhead{Funding}
This work is supported by the European Research Council (ERC) under the Horizon Europe program (Grant-No. 101052785, project: SoftFrac).
\bmhead{Data availability}
The version of the code, the executable, the parameter settings and the result files are available from the corresponding author upon request.
\bmhead{Conflict of interest}
On behalf of all authors, the corresponding author states that there is no conflict of interest.
\bmhead{Ethical approval}
This study does not involve human participants, animal subjects, or ethical concerns requiring institutional approval.
\bmhead{Replication of results}
All the presented methodology is implemented in C++ utilizing the finite element library deal.II \citep{bangerth2007deal, arndt2021deal}. The version of the code, the executable, the parameter settings and the result files are available from the corresponding author upon request.
\bibliography{references}

\end{document}